\documentclass[10pt,aps,prb,showpacs,superscriptaddress,twocolumn,citeautoscript]{revtex4-2}
\usepackage{graphicx}
\usepackage{dcolumn}
\usepackage{bm}

\usepackage[utf8]{inputenc}
\usepackage[T1]{fontenc}
\usepackage{mathptmx}
\usepackage{etoolbox}
\usepackage{xcolor}

\usepackage{times}

\usepackage{graphicx}
\usepackage{bm}        
\usepackage{amssymb}  
\usepackage{amsmath}
\usepackage{gensymb}
\usepackage{tikz}

\newcommand{\beq}{\begin{equation}}
\newcommand{\eeq}{\end{equation}}

\newcommand{\fr}{\bm{r}}
\newcommand{\fR}{\bm{R}}
\newcommand{\fx}{\bm{\hat{x}}}
\newcommand{\fy}{\bm{\hat{y}}}

\newcommand{\subeqs}[1]{\begin{align} #1 \end{align}}

\begin{document}
\title{Bound states in the continuum in circular clusters of scatterers}
\author{Marc Mart\'i-Sabat\'e}
\affiliation{GROC, UJI, Institut de Noves Tecnologies de la Imatge (INIT), Universitat Jaume I, 12071, Castell\'o de la Plana, Spain}
\author{Bahram Djafari-Rouhani}
\affiliation{IEMN, University of Lille, Cité scientifique, 59650 Villeneuve d’Ascq, France}
\author{Dani Torrent}
\email{dtorrent@uji.es}
\affiliation{GROC, UJI, Institut de Noves Tecnologies de la Imatge (INIT), Universitat Jaume I, 12071, Castell\'o de la Plana, Spain}
\date{\today}

\begin{abstract}
In this work, we study the localization of flexural waves in highly symmetric clusters of scatterers. It is shown that when the scatterers are placed regularly in the perimeter of a circumference the quality factor of the resonances strongly increases with the number of scatterers in the cluster. It is also found that in the continuous limit, that is to say, when the number of scatterers tends to infinite, the quality factor is infinite so that the modes belong to the class of the so called bound states in the continuum or BICs, and an analytical expression for the resonant frequency is found. These modes have different multipolar symmetries, and we show that for high multipolar orders the modes tend to localize at the border of the circumference, forming therefore a whishpering gallery mode with an extraordinarily high quality factor. Numerical experiments are performed to check the robustness of these modes under different types of disorder and also to study their excitation from the far field. Although we have focused our study to flexural waves, the methodology presented in this work can be applied to other classical waves, like electromagnetic or acoustic waves, being therefore a promissing approach for the design of high quality resonators based on finite clusters of scatterers. 
\end{abstract}

\maketitle 

\section{Introduction}
Bound states in the continuum (BICs) are eigenmodes of a system whose energy lies in the radiation part of the spectrum while remaining localized in a finite part of the system and with an infinite lifetime. These states were first mathematically proposed in 1929 by von Neumann and Wigner in the framework of quantum mechanics \cite{neumann1929merkwurdige}, although the concept has been extended to classical waves, like acoustics \cite{quotane2018trapped,jin2017tunable,mizuno2019fano,amrani2021experimental,huang2022general}, microwaves \cite{mrabti2020aharonov,mrabti2018transmission} or optics \cite{hsu2013bloch,sadreev2021interference,bulgakov2008bound}. 

Despite the fact that the practical realization of BICs is a challenging problem, structures based on them present sharp resonances with extremely high quality factors, which have as well the advantage, unlike ideal BICs, that can be excited with external radiative fields. Also named quasi-BICs (or QBICs), these modes have been widely used in sensing applications\cite{wu2019giant,yesilkoy2019ultrasensitive,kuhner2022radial}.

Among the wide variety of geometries and structures used to find BICs\cite{hsu2016bound}, those based on finite structures are specially interesting for practical applications, since periodic or waveguide BICs will always present finite-size effects which will decrease their efficiency. For instance, circular clusters of scatterers studied in some recent works\cite{putley2021whispering,kuhner2022radial} are extraordinarily convenient from the practical point of view. In this work, we will generalize the study of these circular clusters of scatterers to provide a general schema for the realization of QBICs based on this geometry.

The manuscript is organized as follows: After this introduction, in section \ref{sec:CircularArray} we study the formation of bound states in the continuum in open systems by attaching a cluster of mass-spring resonators to a thin elastic plate. We will find that when the scatterers in the cluster are arranged in the corners of a regular polygon the quality factor of the resonances quickly increases with the number of scatterers in the cluster. In section \ref{sec:BICRobustness} we perform several numerical experiments to check the robustness of these modes, and in section \ref{sec:ScatteringCrossSection} their excitation from the far field will be considered. Finally, section \ref{sec:summary} sumarizes the work.

\section{Eigenmodes of a polygonal cluster of scatterers}
\label{sec:CircularArray}
The propagation of flexural waves in thin elastic plates where a cluster of $N$ point-like resonators has been attached at positions $\fR_\alpha$ is described by means of the inhomogeneous Kirchhoff\cite{torrent2013elastic} equation
\begin{equation}
(\nabla^4-k_0^4)\psi(\fr)=\sum_{\alpha=1}^Nt_\alpha \delta(\fr-\fR_\alpha)\psi(\fr)
\end{equation}
where $\psi(\fr)$ is the spatial part of the vertical displacement of the plate, which is assumed to be harmonic and of the form
\beq
W(\fr,t)=\psi(\fr)e^{-i\omega t}.
\eeq
Also, the free space wavenumber $k_0$ is given by
\beq
k_0^4=\frac{\rho h}{D}\omega^2,
\eeq
with $\rho$, $h$ and $D$ being the plate's mass density, height and rigidity, respectively. The response of each resonator is given by the $t_\alpha$ coefficient, which is a resonant quantity whose properties depend on the geometry of the scatterer attached to the plate\cite{packo2019inverse}. However, for the purposes of the present work, it will be assumed that it can take any real value in the range $t_\alpha\in (-\infty,\infty)$. 

A self-consistent multiple scattering solution can be found for the above equation as
\beq
\psi(\fr)=\psi_0(\fr)+\sum_{\alpha=1}^N B_\alpha G(\fr-\fR_\alpha)
\eeq
where $\psi_0(\fr)$ is the external incident field on the cluster of scatterers, $G(\fr)$ is the Green's function of Kirchhoff equation,
\beq
G(\fr)=\frac{i}{8k_0^2}(H_0(k_0r)-H_0(ik_0r))
\eeq
with $H_0(\cdot)$ being Hankel's function of first class. The multiple scattering coefficients $B_\alpha$ can be obtained by means of the self-consistent system of equations 
\beq
\label{eq:MST}
\sum_{\beta=1}^NM_{\alpha \beta}B_\beta=\psi(\fR_\alpha),
\eeq
where
\beq
M_{\alpha\beta}=t_{\alpha}^{-1}\delta_{\alpha\beta}-G(\fR_{\alpha\beta})
\eeq
is the multiple scattering matrix $M$. 

The eigenmodes of a cluster of $N$ scatterers attached to a thin elastic plate can be found assuming that there is no incident field, so that the total field excited in the plate is due only to the scattered field by all the particles\cite{marti2021dipolar,marti2021edge}. Under these conditions equation \eqref{eq:MST} becomes a homogeneous system of equations with non-trivial solutions only for those frequencies satisfying 
\beq
\det M(\omega)=0.
\eeq

For finite clusters of scatterers the above condition can be satisfied only for complex frequencies, being the inverse of the imaginary part of this frequency the quality factor of the resonance. Those configurations in which the imaginary part of the resonant frequency is extraordinarily small (hence the quality factor extraordinarily big) receive the name of quasi-BIC or QBIC modes. In the following lines it will be shown that arranging the scatterers in the vertices of regular polygons we can obtain resonances whose quality factor diverges as the number of scatterers approaches to infinite.

Then, if the scatterers are all identical with impedance $t_0$ and they are regularly arranged in a circumference of radius $R_0$ and placed at angular positions $2\pi\alpha/N$, for $\alpha=0,\ldots, N-1$, (as shown in Figure \ref{Fig:Geometry} in Appendix \ref{sec:Appendix})  the Hamiltonian of the system commutes with the rotation operator $R_N$, whose eigenvalues are $\lambda_\ell=\exp(i2\pi \ell/N)$, with $\ell=0,\ldots, N-1$, and this implies a relationship between the coefficients of the form\cite{putley2021whispering}
\begin{equation}
B_\alpha^\ell=e^{2i\pi\ell\alpha/N}B_0^\ell,
\label{eq:Bcoeff}
\end{equation}
thus equation \eqref{eq:MST} becomes 
\begin{equation}
\label{eq:redMST}
(1-t_0\sum_\beta G(\fR_{0\beta})e^{2i\pi\ell\beta/N})B_0^\ell=0.
\end{equation}
It is more suitable now to define the Green's function as
\begin{equation}
G(\fr)\equiv ig_0\xi (\fr)
\label{eq:greenFunction}
\end{equation}
where
\begin{equation}
g_0=\frac{1}{8k_0^2}
\end{equation}
and
\begin{equation}
\xi(\fr)=H_0(k_0r)-H_0(ik_0r),
\end{equation}
so that $\xi(\bm{0})=1$ and $\gamma_0=t_0g_0$ is a real quantity.  The eigenmodes of the system are found as the non-trivial solutions of equation \eqref{eq:redMST}, thus for the $\ell$-th mode we need to solve
\begin{equation}
1-i\gamma_0\sum_\beta \xi(\mathbf{R}_\beta) e^{2i\pi\ell\beta/N}=0.
\end{equation}

This equation will give us a set of complex free-space wavenumbers $k_0^n$ from which we can obtain the eigenfrequencies $\omega_n$ by means of the plate's dispersion relation. The imaginary part of these eigenfrequencies is related with the quality factor of the mode: the lower the imaginary part the larger the quality factor, thus a BIC will be found if we can obtain a real wavenumber $k_0^n$ satisfying the above equation. Thus, assuming this wavenumber exists, we define
\begin{equation}
S^\ell=\sum_\beta \xi_\beta e^{2i\pi\ell\beta/N}=S_R^\ell+iS_I^\ell,
\label{eq:Sl}
\end{equation}
and the secular equation can be divided in real and imaginary parts as
\subeqs{
	S_R^\ell(k_0)&=0\label{eq:SR}\\
	1+\gamma_0S_I^\ell(k_0)&=0.\label{eq:SI}
}
The second of these equations will always be satisfied, since $\gamma_0$ is a resonant factor that can be selected to run from $-\infty$ to $\infty$. Therefore, we have to find the conditions for which the first of the equations can be satisfied.

\begin{figure}[h!]
	\includegraphics[width = \columnwidth]{ 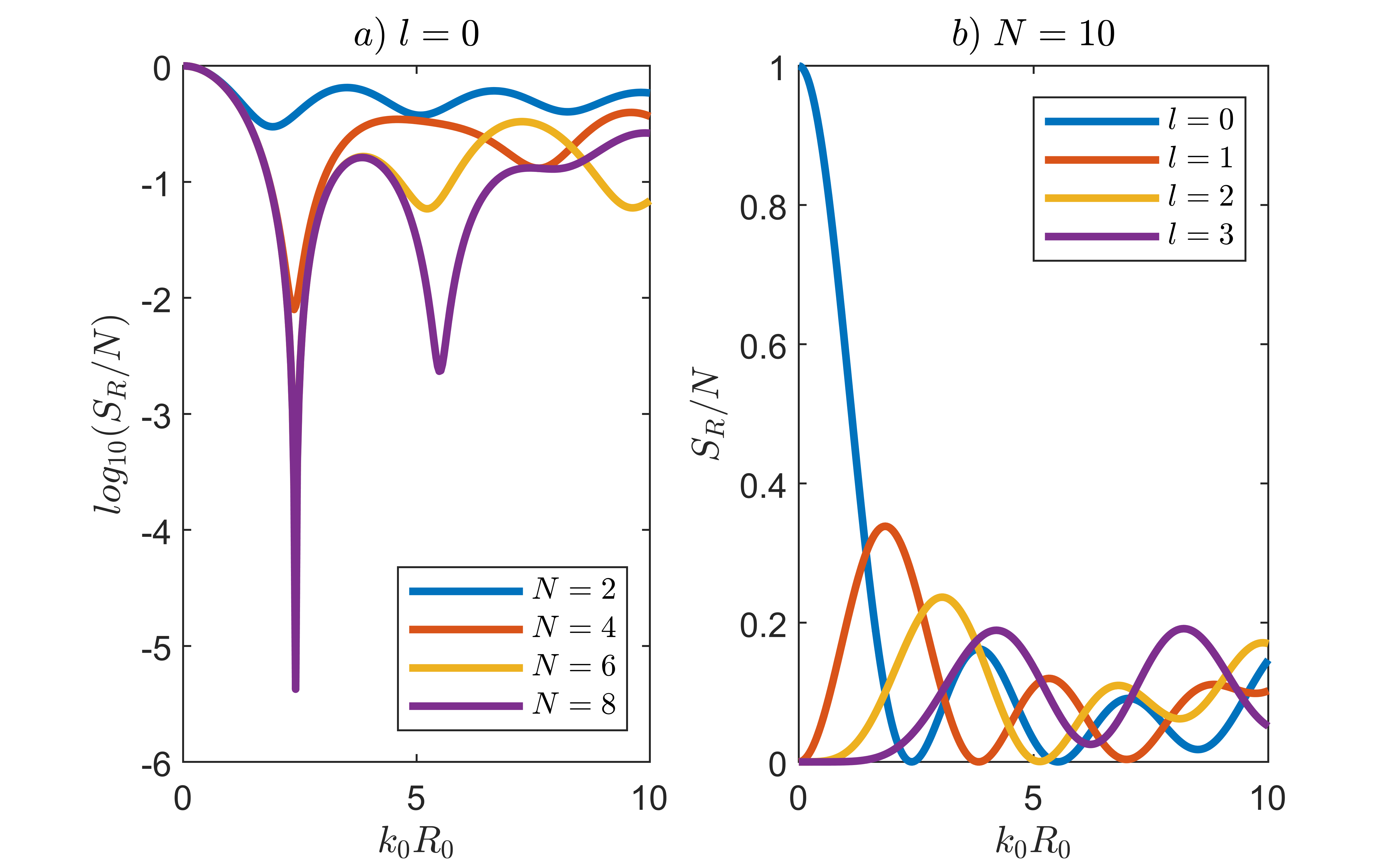}
	\caption{\label{Fig:SREvolution}{$S_R$ summation for different situations. In panel $a$ the different lines correspond to different number of scatterers in the cluster, and the resonance index is fixed at $l = 0$. In panel $b$, the number of scatterers in the cluster is fixed ($N = 10$) and the evolution of $S_R/N$ as a function of $k_0$ is shown for different resonant index.}}
\end{figure}

Figure \ref{Fig:SREvolution}, panel $a$, shows the evolution of $S_R^\ell$ (in logarithmic scale, for clarity) as a function of $k_0R_0$ for $\ell=0$ and for different number of scatterers $N$ in the cluster. As can be seen, for a small number of scatterers the function does not approach zero, so that no BIC can be found, although for a relatively large number of particles the function is nearly zero indicating a high-quality resonance. Panel $b$ in figure \ref{Fig:SREvolution} shows $S_R^\ell$ as a function of $k_0R_0$ but for a fixed number of scatterers $N = 10$ and for $\ell=0,1, 2, 3$. In this case, we can see how the function $S_R^\ell$ is nearly zero for low $\ell$, although for $\ell=3$ the minimum is actually far away the zero value. It is found numerically that these minima approach to zero as we increase the number of scatterers in the cluster, although the zero value is reached only in the limit $N\to\infty$, indeed it is found that (see Appendix \ref{sec:Appendix} )
\begin{equation}
	\lim_{N\to\infty}\frac{1}{N}S_R^\ell=J_\ell^2(k_0R_0),
	\label{eq:LimEq}
\end{equation}
consequently the resonances of the cluster are given by the zeros of the Bessel function $J_\ell(k_0R_0)$ in this limit, reaching the BIC condition, although  in clusters with $N>10$ good quality resonances are found, being therefore quasi-BIC modes. It is interesting to mention that the position of the resonances is independent of the number of particles $N$, although the corresponding impedance $\gamma_0$ has to be obtained from equation \eqref{eq:SI} which will be, in general, a function of $N$.

\begin{figure}[h!]
	\includegraphics[width=\columnwidth]{ 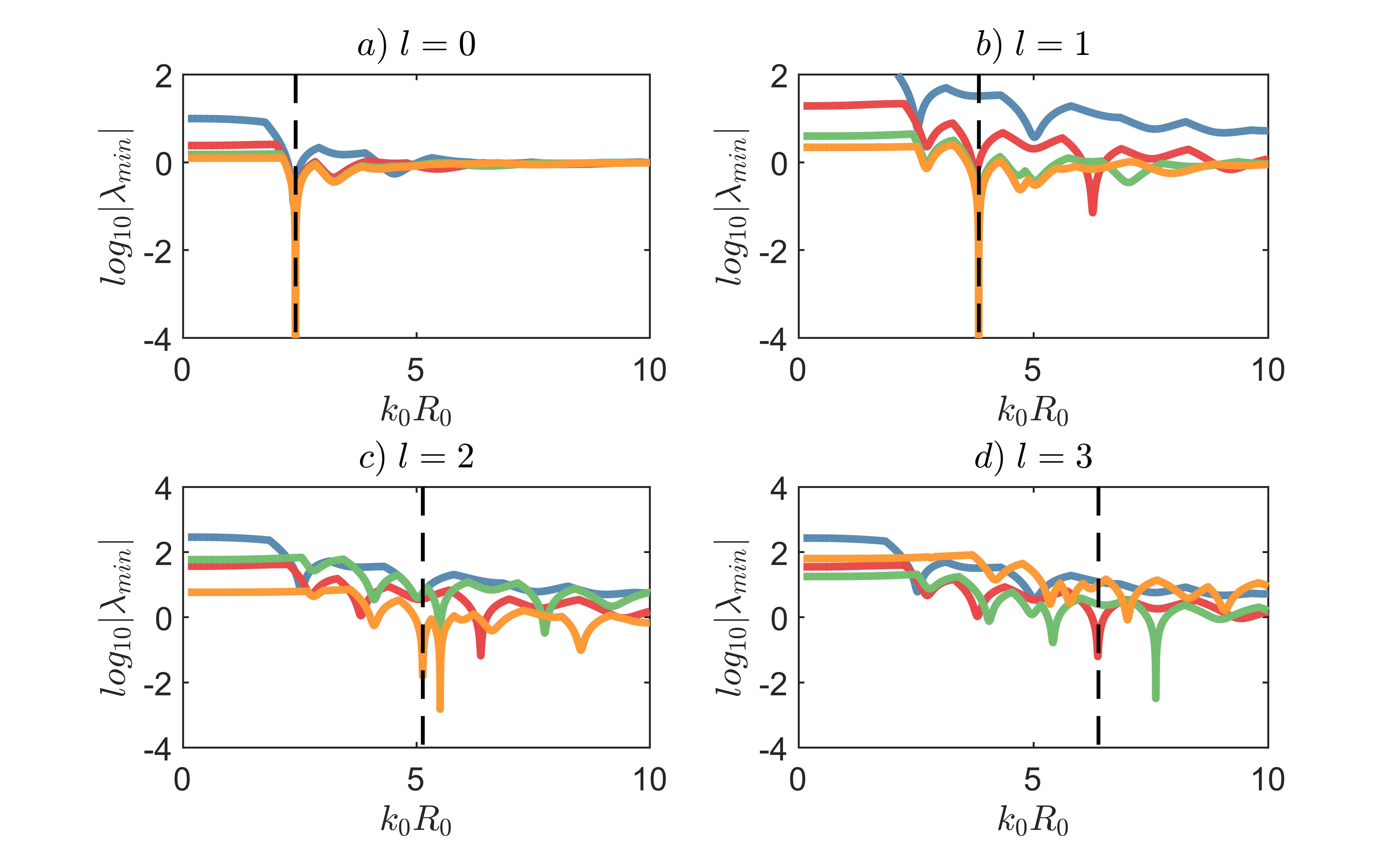}
	\caption{\label{Fig:lastEigenvalueComparison}{Resonance comparison for several clusters. Each panel presents the resonances for a different resonant index ($\ell$). The colour code is the same for the four panels, representing a different number of scatterers in the cluster (blue is $N = 4$, red is $N = 6$, green is $N = 8$ and orange is $N = 10$). The dashed line indicates the frequency at which the resonance is predicted for an infinite number of scatterers in the cluster.}}
\end{figure}

\begin{figure}[h!]
	\includegraphics[width=\columnwidth]{ 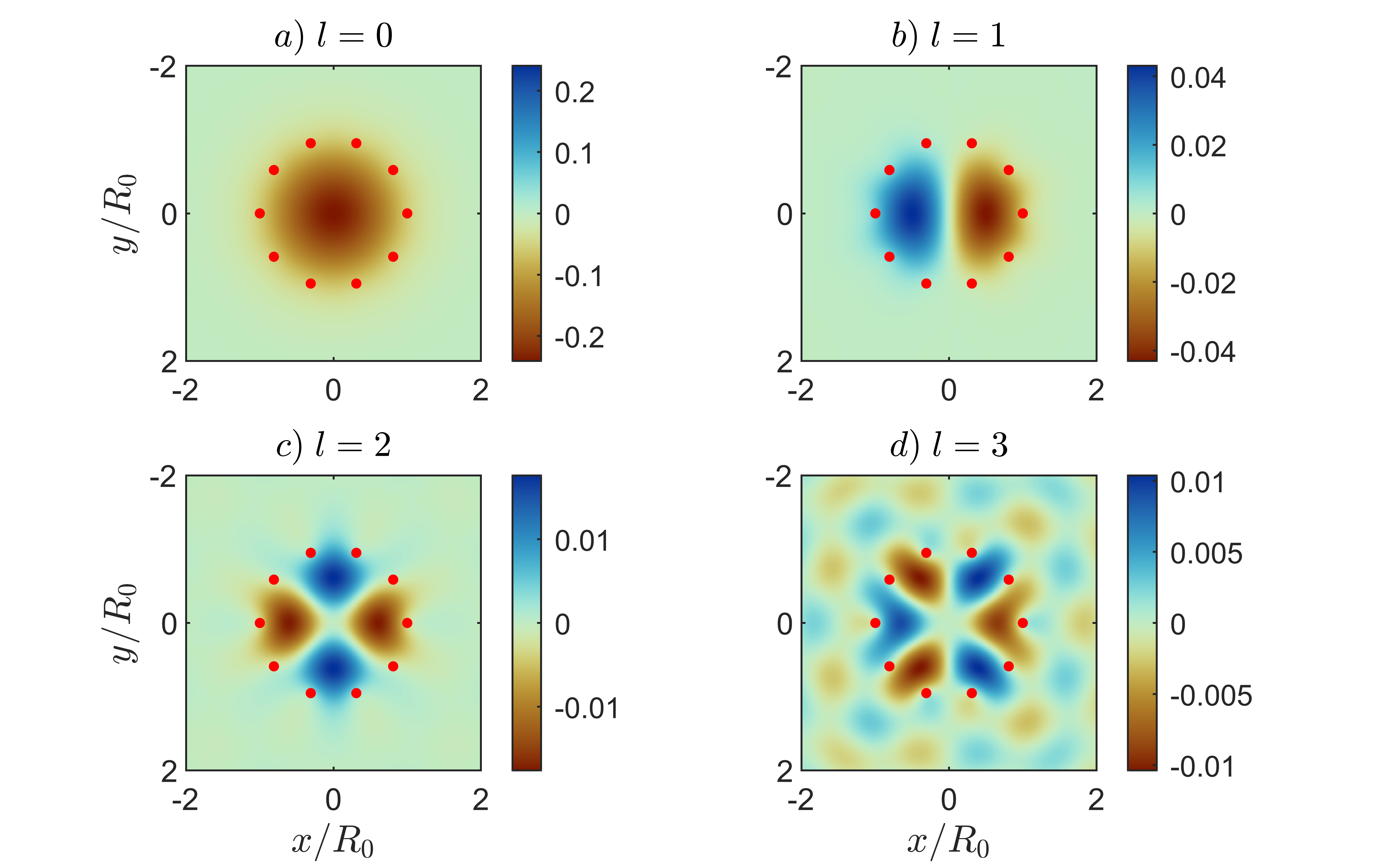}
	\caption{\label{Fig:InnerField}{Real part of the eigenfunction for different resonant index. The clusters have the same number of scatterers ($N = 10$).}}
\end{figure}

The quality factor of these resonances can be found by the analysis of the minimum eigenvalue of the multiple scattering matrix $M$\cite{marti2021edge,doi:10.1063/5.0098239}. Figure \ref{Fig:lastEigenvalueComparison}, panels $a$, $b$, $c$ and $d$, show this parameter for the modes $\ell=0,1,2,3$, respectively. Results in each plot are shown for clusters of $N=4,6,8$ and $10$ particles, and it is clearly seen how the quality factor of the resonance increases with the number of scatterers. The vertical dashed line is the frequency at which the function in equation \eqref{eq:LimEq} cancels, that is to say, the frequency at which the resonance is predicted for a cluster with an infinite number of scatterers. When higher resonances are studied, some resonances disappear for the smaller clusters. This is the case of $\ell = 2$ (panel $d$), where the resonance only appears for $N = 8$ and $N = 10$. Something remarkable happens in the $\ell = 3$ case; the resonance is present in the $N = 6$ cluster, whereas the rest of the clusters do not present any resonance. As can be seen in figure \ref{Fig:InnerField}, $\ell = 3$ shows a $\pi / 3$ symmetry in the inner field. In fact, the resonant index $\ell$ defines the symmetry of the eigenmode as $\pi/\ell$. Thus, it is easier to excite this resonance when the number of scatterers is a multiple of the symmetry of the mode. It is worth mentioning that other modes appear in this analysis given that we are plotting the full multiple scattering matrix $M$, without any hypothesis on the symmetry of the mode, therefore all the multipolar resonances will result in minima in the determinant of $M$.

Figure \ref{Fig:InnerField} shows the corresponding eigenfunctions for the largest cluster ($N = 10$), showing how the index $\ell$ defines the symmetry of the mode. It is also noticeable how as long as the $\ell$ index increases, the eigenfunction is less confined inside the cluster. This is a direct consequence of the decrease of the quality factor of the resonance and the leakage of energy into the bulk.

\begin{figure}[h!]
	\includegraphics[width=\columnwidth]{ 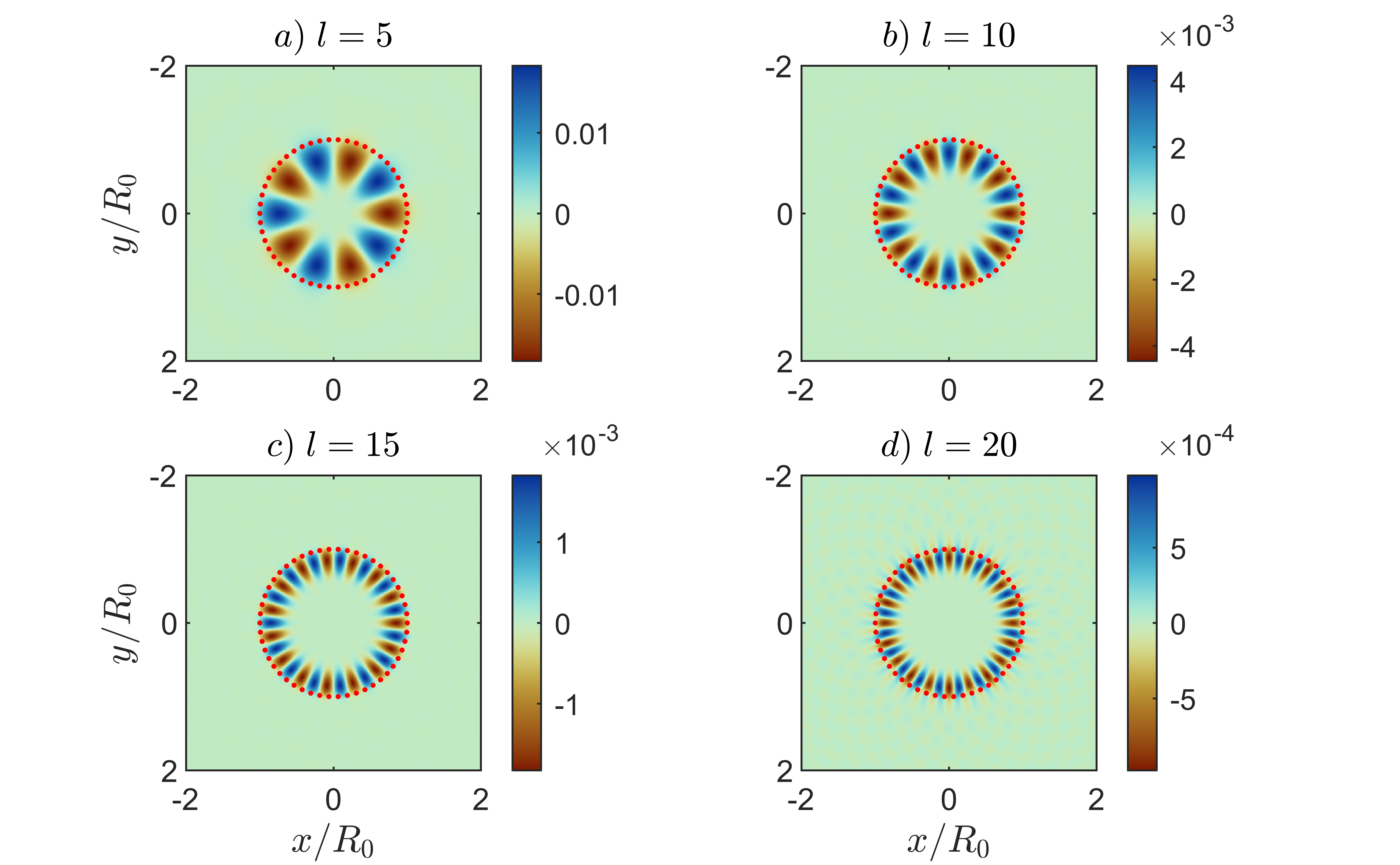}
	\caption{\label{Fig:InnerFieldV2}{Real part of the eigenfunction for different resonant index. The clusters have the same number of scatterers ($N = 50$).}}
\end{figure}

Modes of high index $\ell$ tend to localize near the scatterers, resulting in the so-called whispering  gallery modes. This approach allows therefore for the systematic design of high-quality whispering gallery modes. Figure \ref{Fig:InnerFieldV2} shows examples of these modes for a cluster of $N=50$ scatterers and indexes $\ell=5,10,15,20$. The localization of the field near the perimeter of the cluster as we increase $\ell$ is evident in these plots.

\section{Robustness of the quasi-BIC modes}
\label{sec:BICRobustness}
In this section, several numerical simulations are presented, which objective is to study how the modes get deformed or destroyed when the positions of the scatterers in the cluster are perturbed.

\begin{figure}[h!]
	\includegraphics[width=\columnwidth]{ 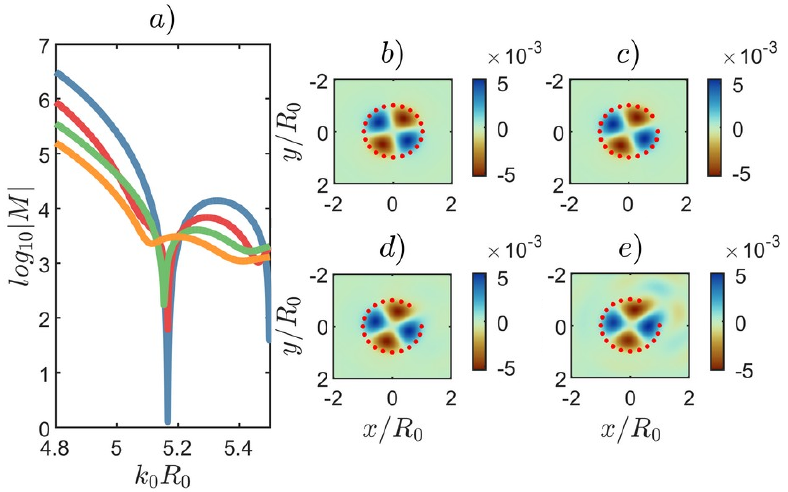}
	\caption{\label{Fig:ClusterOpening}{Disappearence of the BIC resonance when some scatterers are missing in the circular array. At left, the evolution of the resonance; the blue line represents the cluster with all the scatterers present, in the red one one scatterer is missing, the green line is for two missing scatterers, and the orange line is for three missing scatterers. The total number of resonators is 20. The resonance index is $\ell = 2$. At right, both maps show the eigenfunctions (real value) for the original situation and the three times deformed cluster.} }
\end{figure}

The first situation considers missing scatterers in the polygonal arrangement. Figure \ref{Fig:ClusterOpening}, panel $a$, shows the plot of the minimum eigenvalue of the multiple scattering matrix as a function of frequency when all the scatterers are present (blue line), and then when we remove one (red), two (green) or three (orange) adjacent scatterers. The total number of resonators in the cluster is $N=20$, and the explored resonance is $\ell = 2$. We see how frequency of the resonance is slightly displaced and its quality factor decreases. The quality factor of the original resonance is $Q = 1880$; $Q = 1086$ after deleting one scatterer, $Q = 392$ after deleting the second one and the resonance disappears when the third resonator is removed. 

Panels $b$ to $e$ of figure \ref{Fig:ClusterOpening} show the maps of the mode for the different situations described above. It is clear that the symmetry of the mode is generally preserved and the field is still localized inside the cluster, although the leakage is strong when three scatterers are removed from the cluster, as can be understood from the broadening of the peak shown in the panel $a$.

From the practical point of view it is also interesting to analyze the quality of the resonances with positional disorder of the particles in the cluster, since this is something we cannot avoid in practical realizations of these structures. Then, the positional disorder has been applied to each scatterer in its angular position, such that 
\begin{equation}
\theta_\beta= 2\pi \frac{\beta}{N} + \sigma \mathcal{N}(0,1),
\end{equation}
where $\mathcal{N}(0,1)$ is a normal distribution of zero mean and unitary variance, and $\sigma$ is the variance of the disorder we aim to apply. Therefore, all the scatterers remain in the same circle of radius $R_0$, but they are no longer equally distributed all along it.

\begin{figure}[h!]
	\includegraphics[width=\columnwidth]{ 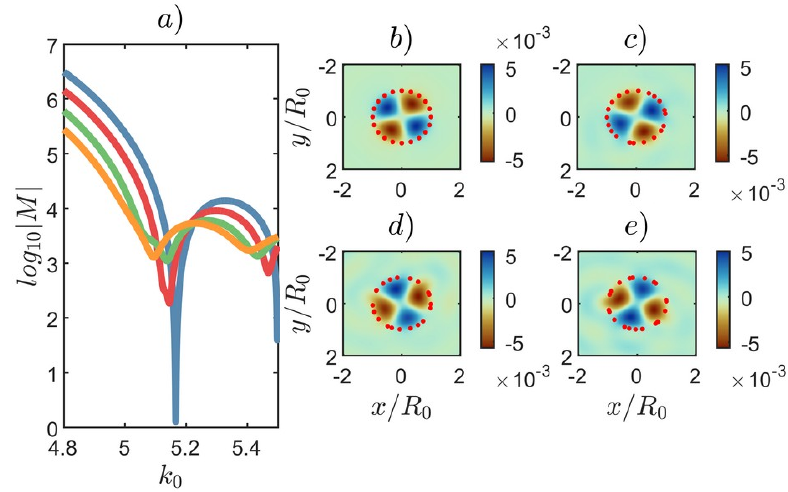} 
	\caption{\label{Fig:PositionalDisorder}{Disappearence of the BIC resonance the position of the resonators is slightly changed. At left, the evolution of the resonance; the blue line represents the cluster at the original configuration, the red, green and orange lines show the resonance with increasing percentage of disorder in the position of the scatterers. The maps at right show the eigenfunctions (real value) for the four configurations.} }
\end{figure}

Figure \ref{Fig:PositionalDisorder} shows the same results as figure \ref{Fig:ClusterOpening} but for the positional disorder just described, with $\sigma=5\times \pi/180$ for the red line, $7.5\times \pi/180$ for the green one and $10\times \pi/180$ for the orange one. We see how the quality factor of the resonance is strongly reduced as the disorder is increased, although the quadrupolar symmetry of the mode still remains. 

These results show that, although the quality factor of the resonances is strongly sensitive to the perturbations of the cluster, their symmetry is a robust parameter against disorder. We have also seen that the frequency of the resonance is weakly disturbed.

\section{Excitation of quasi-BICs from the continuum}
\label{sec:ScatteringCrossSection}
In this section we will explore the possibility of exciting and detecting quasi-BICs by means of external incident fields to the cluster. The excitation of BICs by means of incident plane waves is impossible, since these states belong to the continuum and BICs do not couple to them. However, quasi-BICs can in principle be excited by these fields resulting in strong peaks in the scattering cross section of the cluster, which can be used for instance for sensing applications.
 
\begin{figure}[h!]
	\includegraphics[width=\columnwidth]{ 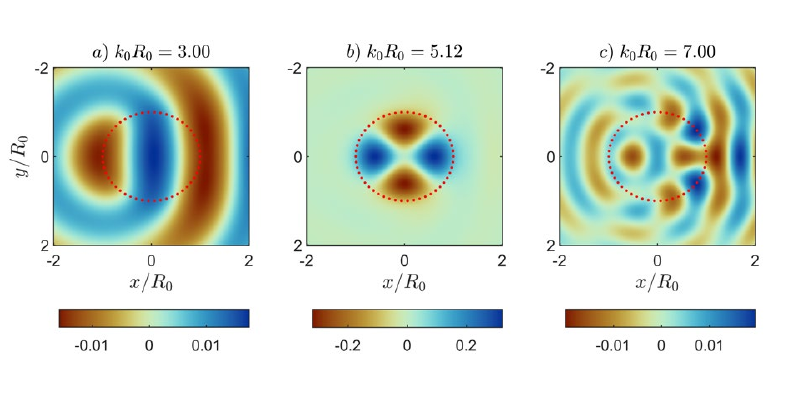}
	\caption{\label{Fig:InnerMaps}{Sccatered field from a $N = 50$ resonators' cluster for three different frequencies. The bound state in the continuum is predicted to happen at the second frequency ($k_0 R_0 = 5.118$). While the elastic field is completely located inside the circle in the middle panel, both right and left panels show the energy distributed all along the plate. The maximum displacement field is bigger in the middle panel than in the other two.} }
\end{figure}

Figure \ref{Fig:InnerMaps}  shows an example of the scattered field by a cluster of $N=50$ scatterers when a plane wave propagates along the $x$ axis. Simulations are shown for three different wavenumbers. Panels $a$ and $c$ show non-resonant frequencies, while the panel $b$ shows the scattered field at the quasi-BIC condition, showing how, although some scattered field leaves the cluster, most of the scattering energy remains confined inside it.

The analysis of the excitation of a quasi-BIC mode can be done by means of the far field radiated by the cluster upon plane wave incidence at frequencies near the quasi-BIC condition. The far-field radiation function is given by
\begin{equation}
f(\theta) = \sum_{\beta = 1}^{N} B_{\beta} e^{-ik_0R_\beta \cos{(\theta - \theta_{\beta})}},
\end{equation}
and the total scattering cross-section $\sigma_{sca}$ is computed as\cite{packo2021metaclusters}
\begin{equation}
\sigma_{sca} = \frac{1}{16\pi D k_0^2} \int_0^{2\pi}|f(\theta)|^2 d\theta.
\end{equation}

\begin{figure}[h!]
	\includegraphics[width=\columnwidth]{ 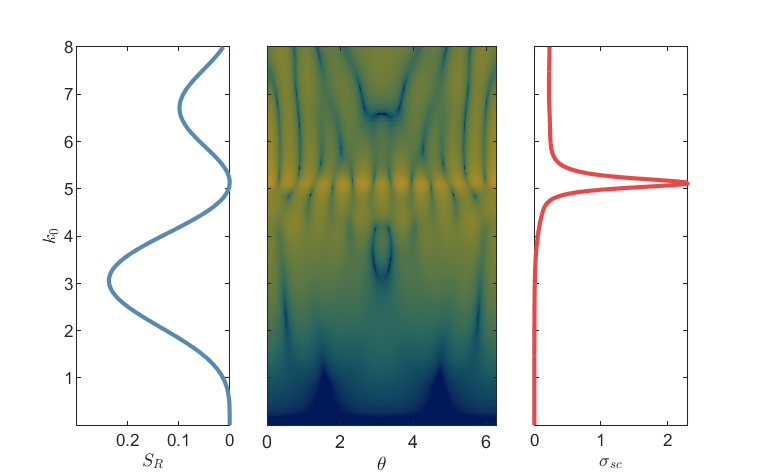}
	\caption{\label{Fig:FarField}{Far-field radiation pattern and scattering cross-section. The left graph represents the $S_R$ term as a function of the frequency of the system. The central map shows the far-field radiation pattern ($f(\theta)$) as a function of the angle and the frequency. Finally, the right graph represents the scattering cross-section as a function of the frequency. As it can be seen, the zero of the $S_R$ summation term finds a peak in both the far-field radiation pattern and the scattering cross-section.} }
\end{figure}

Figure \ref{Fig:FarField} shows the far-field analysis for the example shown in figure \ref{Fig:InnerMaps}. The left panel shows the function $S_R^2$, showing the minima where the resonance is expected ($k_0 R_0=5.118$).  We can see how at this frequency there is an enhancement of the far-field pattern $f(k_0,\theta)$ shown in the central panel, although the symmetry of this radiation pattern does not corresponds to that of the quasi-BIC mode. The reason is that the mode is confined inside the cluster, thus the $\ell=2$ symmetry can be observed only in the near field, but this field interacts with the $N=50$ scatterers of the cluster and excite some radiation far field with a multipolar symmetry. The right panel shows how the total scattering cross section $  \sigma_{sca} $ is enhanced at the resonant condition, as expected.

\section{Summary}
\label{sec:summary}
In summary, we have studied the possibility of having bound states in the continuum (BICs) in clusters of scatterers for flexural waves in thin plates. We found that a polygonal arrangement, which would become a circular scatterer when the number of scatterers tends to infinite, presents resonances of divergent quality factor, thus these modes can be defined as quasi-BIC modes. We also derived an analytical expression for the resonant frequency of the different multipolar resonances of the circular scatter which is accurate as well for finite clusters. Several numerical experiments show that these modes are robust in general, in the sense that only the quality factor is significantly changed when different types of disorder are applied, while the resonant frequency is only weakly distorted. We found as well that the quasi-BIC modes can be excited from the continuum, since a peak in the total scattering cross section is detected, which enhances the possible applications of these structures for sensing applications. The formulation based on multiple scattering theory shows as well that this approach is not unique of flexural waves but it could also be applied to other type of classical or quantum waves, with similar results expected.





\section*{Acknowledgments}
D.T. acknowledges financial support through the ``Ram\'on y Cajal'' fellowship, under Grant No. RYC-2016-21188. Research supported by DYNAMO project (101046489), funded by the European Union. Views and opinions expressed are however those of the authors only and do not necessarily reflect those of the European Union or European Innovation Council. Neither the European Union nor the granting authority can be held responsible for them. Marc Mart\'i-Sabat\'e acknowledges financial support through the FPU program under grant number FPU18/02725.
\appendix
\section{Appendix: Continuous limit of the cluster's Green's function}
\label{sec:Appendix}
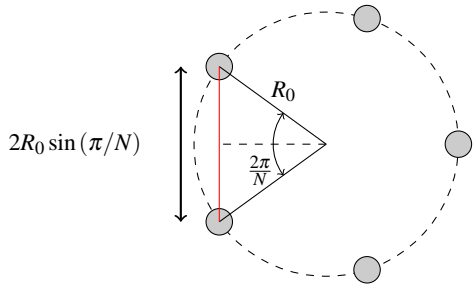
\begin{figure}[h!]
	\centering
	\begin{tikzpicture}
		
		\draw[dashed] (0,0) circle (50pt);
		\draw[fill=black!20!white] (50pt,0) circle (5pt);
		\draw[fill=black!20!white] (15.45pt,47.55pt) circle (5pt);
		\draw[fill=black!20!white] (-40.45pt,29.39pt) circle (5pt);
		\draw[fill=black!20!white] (-40.45pt,-29.39pt) circle (5pt);
		\draw[fill=black!20!white] (15.45pt,-47.55pt) circle (5pt);
		
		\draw (0,0) -- (-40.45pt,29.39pt);
		\draw (0,0) -- (-40.45pt,-29.39pt);
		\draw[red] (-40.45pt,29.39pt) -- (-40.45pt,-29.39pt);
		
		\draw[dashed] (0,0) -- (-40.45pt,0);
		
		\draw (-16pt,20pt) node {$R_0$};
		\draw[<->] (-16.18pt,11.76pt) arc (144:216:20pt);
		\draw (-25pt,-10pt) node {$\frac{2\pi}{N}$};
		
		\draw [<->,thick] (-55pt,29.39pt) -- (-55pt,-29.39pt);
		\draw (-95pt,0pt) node {$2R_0\sin{(\pi/N)}$};
		
	\end{tikzpicture}
	\caption{\label{Fig:Geometry}{Cluster's geometry.} }
\end{figure}

In this appendix we will derive an analytical expression for the sum $S_R^\ell$ when the number of scatterers in the circular array tends to infinite. According to figure \ref{Fig:Geometry}, the scatterers in the cluster are placed in the vertices of a regular polygon of $N$ sides, thus the position of the $\alpha$ scatterer is given by
\beq
\fR_\alpha=R_0\cos2\pi/N\alpha \fx+R_0\sin2\pi/N\alpha \fy
\eeq
In the limit of  $N \longrightarrow \infty$, the variable $\theta_\alpha=2 \pi \alpha / N$ can be substituted by a continuous variable $\theta \in [0,2\pi]$, such that $d\theta=2\pi/N$. Also, the distance $R_{0\alpha}$ between the scatterer of reference and any scatterer in the cluster is, according to figure \ref{Fig:Geometry},
\beq
R_{0\alpha}=2R_0\sin \frac{\pi}{N}
\eeq
which, in the limit $N \longrightarrow \infty$ becomes
\beq
R(\theta)=2R_0\sin \frac{\theta}{2}
\eeq
Thus, we can write
\begin{equation}
\lim_{N\to\infty}\frac{1}{N}S_R^\ell= \frac{1}{2\pi}\text{Re}{\int_0^{2\pi} \xi(\theta) e^{i \ell \theta}d\theta}.
\end{equation}
with
\begin{equation}
\xi(\theta) = H_0(k_0R(\theta)) + \frac{2i}{\pi}K_0(k_0R(\theta)),
\end{equation}

For $\ell = 0$ we have
\begin{align*}
	\lim_{N\to\infty}\frac{1}{N}S_R^0 = \frac{1}{2\pi}\int_0^{2\pi} J_0(2k_0R_0\sin{(\theta/2)})d\theta \\
	= \frac{2}{\pi}\int_0^{\pi/2}J_0(2k_0R_0\sin{\theta})d\theta.
\end{align*}

By using the following identity \cite{gradshteyn2014table}

\begin{equation}
\int_0^{\pi/2}J_{2\nu}(2z\sin{x})dx = \frac{\pi}{2}J_\nu^2(z),
\end{equation}
we arrive to
\begin{equation}
\lim_{N\to\infty}\frac{1}{N}S_R^0 = J_0^2(k_0R_0).
\end{equation}

Similarly, for $\ell \neq 0$, we have now
\begin{equation}
\lim_{N\to\infty}\frac{1}{N}S_R^\ell= \frac{1}{2\pi}\Re{\int_0^{2\pi} \chi_\theta e^{ i l \theta}d\theta}
\end{equation}
thus
\begin{align*}
	\frac{1}{2\pi}\int_0^{2\pi} J_0(2k_0R_0\sin{\theta/2})e^{i l \theta}d\theta \\
	= \frac{1}{(2\pi)^2}\int_0^{2\pi} \int_{-\pi}^\pi e^{-i2k_0R_0\sin{(\theta/2)}\sin{\tau}}e^{ i l \theta}d\tau d\theta\\
	= \frac{1}{2\pi^2}\int_0^{\pi} \int_{-\pi}^\pi e^{-i2k_0R_0\sin{(\theta)}\sin{\tau}}e^{2 i l \theta}d\tau d\theta\\
	= \frac{(-1)^{2l}}{2\pi}\int_{-\pi}^{\pi} J_{2l}(2k_0R_0\sin{\tau})d\tau \\
	= \frac{2(-1)^{2l}}{\pi} \int_0^{\pi/2} J_{2l}(2k_0R_0\sin{\tau})d\tau
\end{align*}
so that we have
\begin{equation}
\lim_{N\to\infty}\frac{1}{N}S_R^\ell =  J_{\ell}^2(k_0R)
\end{equation}

\end{document}